# ADVERSARIAL-PLAYGROUND: A Visualization Suite for Adversarial Sample Generation

Andrew Norton and Yanjun Qi
*Department of Computer Science*
*University of Virginia*
*Charlottesville, VA 22904, USA*
{apn4za, yanjun}@virginia.edu

## Abstract

With growing interest in adversarial machine learning, it is important for practitioners and users of machine learning to understand how their models may be attacked. We present a web-based visualization tool, ADVERSARIAL-PLAYGROUND, to demonstrate the efficacy of common adversarial methods against a convolutional neural network. ADVERSARIAL-PLAYGROUND provides users an efficient and effective experience in exploring algorithms for generating adversarial examples — samples crafted by an adversary to fool a machine learning system. To enable fast and accurate responses to users, our webapp employs two key features: (1) We split the visualization and evasive sample generation duties between client and server while minimizing the transferred data. (2) We introduce a variant of the Jacobian Saliency Map Approach that is faster and yet maintains a comparable evasion rate [1].

## 1 Introduction

Deep Neural Networks (DNNs) and Convolutional Neural Networks (CNNs) are an essential tool for many machine learning tasks, especially image classification [6]. Unfortunately, recent studies of *evasive samples* show that intelligent attackers can force these models to misclassify samples by performing nearly imperceptible modifications to the sample before attempting classification [5, 10]. These samples are usually crafted through an optimization procedure that searches for small and effective perturbations (details in Section 2). We introduce a visualization tool that enables deep learning practitioners and users to understand how their classifiers may be attacked.

Investigating the behavior of machine learning systems in adversarial environments is an emerging topic at the junction of machine learning and computer security [2]. While machine learning models may appear to be effective for many security tasks like malware classification [7] and face recognition [9], it is important to realize that these classification techniques were not designed to withstand manipulations made by intelligent and adaptive adversaries. In contrast with applications of machine learning to other fields, security tasks involve adversaries that may respond maliciously to the classifier [2]. Therefore, we seek to provide an educational tool for visualizing adversarial examples generated by common evasion attacks and showing how these examples fool a state-of-the-art CNN model.

Our proposed package follows the spirit of *TensorFlow Playground* — a web-based educational tool that helps users understand how neural networks work [11]. TensorFlow Playground has been used in many classes as a pedagogical aid and helps the self-guided student learn more. Its impact inspires us to visualize adversarial samples with a similar tool, which we call "ADVERSARIAL-PLAYGROUND." This web-based visualization tool assists users in understanding and comparing the impact of standard evasion techniques on deep learning models. ADVERSARIAL-PLAYGROUND provides quick and effective visualizations of adversarial examples through two key methods:

**Client/Server Framework** Launching ADVERSARIAL-PLAYGROUND starts a lightweight Python webserver that hosts a collection of pages for the visualizations. Using a server-based approach allows remote hosting on a powerful machine, but the tool may be hosted on any computer running TensorFlow. To improve response time, the server does not transmit large images of evasive samples to the client, sending only a matrix describing the sample and the vector of classifications; the graphics are rendered on the client-side using JavaScript. This design differs from TensorFlow Playground, which solely used client-side

---

[1] Project source code and data from our experiments is available at https://github.com/QData/AdversarialDNN-Playground.

technology with no server-side computations.

**Modified JSMA** We introduce a new, faster variant of the Jacobian Saliency Map Approach (JSMA) that maintains a comparable evasion rate to the original. Most state-of-the-art evasion algorithms are slow due to expensive optimization and the large feature space involved in image classification [3, 4]. For studies focusing only on effectiveness of attack, this is not a major issue. In our application, however, slow visualizations will negatively impact users. Instead of performing a costly search of all feature pairs (as in JSMA), we use a heuristic approximation to reduce the search space considerably. Experiments verify that our faster JSMA maintains the same evasion rate as the usual JSMA algorithm, but executes almost twice as fast.

The rest of this paper takes the following structure: Section 2 discusses three types of state-of-the-art evasion algorithms, Section 3 introduces the system organization and software design of ADVERSARIAL-PLAYGROUND, Section 4 presents the altered JSMA algorithm with an empirical comparison to `cleverhans` JSMA, and Section 5 concludes the paper by discussing possible extensions.

## 2 Background of Adversarial Examples

Studies regarding the behavior of machine learning models in adversarial environments generally fall into one of three categories: (1) *poisoning attacks*, in which specially crafted samples are injected into the training of a learning model, (2) *privacy-aware methods*, which aim to preserve the privacy of information in training data, or (3) *evasion attacks*, in which the adversary aims to create inputs that are misclassified by a target classifier. Our work focuses on evasion attacks.

The goal of evasion is to craft an input for a particular classifier that, while improperly classified, reveals only slight alteration to a human viewer. To formalize the extent of allowed alteration, evasion algorithms minimize the difference between the "seed" image and the resulting evasive sample based on a selected norm (distance function).

In some cases, the adversary specifies the "target" class of an evading sample — for example, the adversary may desire an image that looks like a "5" to be classified as a "7". This is referred to as a *targeted* approach. Conversely, if the adversary does not specify the desired class, the algorithm is considered to be *untargeted*.

Formally, let us denote $f : X \to C$ to be a classifier that maps the set of all possible inputs, $X$, to some finite set of classes, $C$. Then, given a target class $y_t \in C$, a starting sample $x \in X$, and a norm $\|\cdot\|$, the goal of targeted adversarial sample generation is to find $x' \in X$ such that:

$$x' = \arg\min_{s \in X} \{\|x - s\| : f(s) = y_t\} \quad (1)$$

Similarly, in the untargeted case, the goal is to find $x'$ such that:

$$x' = \arg\min_{s \in X} \{\|x - s\| : f(s) \neq f(x)\} \quad (2)$$

In this formalization, we see there are two key degrees of freedom in creating a new evasion algorithm: targeted vs. untargeted attacks and the choice of norm. The latter category provides a useful classification scheme for evasion algorithms, suggested by Carlini and Wagner [3].

### 2.1 $L^0$ Norm

A simple way to determine the extent of the difference between two images is to count the number of pixels that differ between them. That is, if $x$ is our original image, and $x' = x + r$ is the evading image (for some suitable value of $r$), then we can compute the $L^0$ distance between $x$ and $x'$ as following, where $[\cdot]$ is the Iverson bracket[2]:

$$\|r\|_0 = \sum_i [r_i \neq 0] \quad (3)$$

Papernot, et al., suggested using the $L^0$ norm for evaluating the similarity of the initial sample and adversarial result [8]. Their approach computes a saliency map of a given input, ranking pixels based on their contribution to classification. Then, they perform a combinatorial search over all pixel pairs to find the optimal two pixels to adjust. This is repeated until either the modified image is misclassified or the $L^0$ distance between the modified and unmodified image is exceeds a threshold.

This algorithm — as well as the $L^\infty$ fast gradient sign method — was included in the `cleverhans` package by Goodfellow, Papernot, and McDaniel [4]. The specifics of this algorithm will be discussed in more detail in Section 4.1.

### 2.2 $L^2$ Norm

A disadvantage of the $L^0$ norm is that it is not differentiable. Since the $L^2$ norm is differentiable, it is more easily understood from a theoretical standpoint. This norm measures the standard Euclidean distance between two vectors; using the same notation as before, with $x$ as the

---

[2]The Iverson bracket is defined as follows: $[P] = \begin{cases} 1 & P \text{ is true} \\ 0 & \text{otherwise} \end{cases}$

starting vector and $x' = x + r$ for the adversarial input, this norm is computed by:

$$\|r\|_2 = \left(\sum_i r_i^2\right)^{1/2} \quad (4)$$

In their foundational paper on evasive sample generation, Szegedy, et al., posed the issue as a convex optimization problem using the $L^2$ norm [10]. This problem was then solved using the usual (albeit slow) method of box-constrained L-BFGS.

### 2.3 $L^\infty$ Norm

A third commonly used norm is the $L^\infty$ norm, also called the *Chebyshev distance*. This measures the maximal change between two vectors along any single feature. That is, if $x$ is the starting vector and $x' = x + r$ is the adversarial input, the distance between them is computed by:

$$\|r\|_\infty = \max_i \{|r_i|\} \quad (5)$$

The Fast Gradient Sign (FGS) method uses the $L^\infty$ norm in generating evading inputs; this method is commonly used due to its speed [5]. Unlike most prior approaches, which require iteratively changing the evasive sample away from its source class, FGS performs exactly one update step to obtain the evasive input. Informally, the algorithm performs one step of gradient descent, but (1) moves *away* from the loss function's minimum and (2) modifying the input sample rather than the model weights.

If $x$ is our original input, $J(\theta, x, y)$ is the cost function for training the network, and $x'$ is the evading input created by FGS, we have:

$$x' = x + \varepsilon \operatorname{sign}(\nabla_x J(\theta, x, y)) \quad (6)$$

The *attacking power*, $\varepsilon$, may be adjusted to fit the particular domain. Increasing $\varepsilon$ increases likelihood that the evading sample is misclassified, but it also increases $L^\infty$ distance between $x$ and $x'$. This presentation of FGS is untargeted; this algorithm only cares about getting further away from the source class but does not specify any "target" class.

## 3 System Organization: Webserver with Client-side Visualization

As a web application, ADVERSARIAL-PLAYGROUND splits the duties of visualization and computation between the client and server. Through the client, the user adjusts hyperparameters and submits an AJAX request for a generated sample to the server. Once the TensorFlow backend generates the adversarial image and classification likelihoods, the server returns this data to the client. Finally, this information is displayed graphically to the user through use of the `Plotly` JavaScript library (Figure 1).

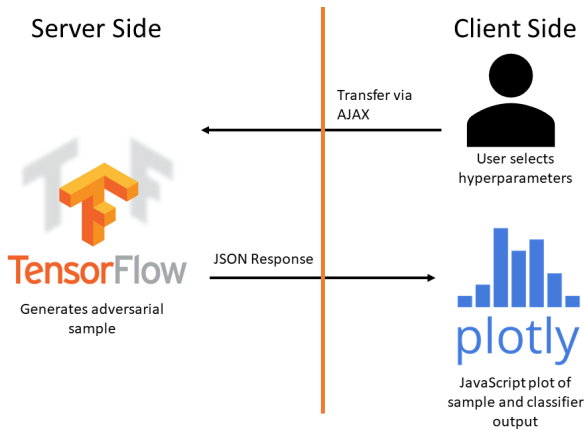

Figure 1: System Interaction

When the application is accessed, the user is presented with the choice between several attack methods — some targeted and some untargeted. After selecting an attack, the user may adjust model parameters, choose the seed image, and specify a target class (if applicable). When the user submits their desired parameters, the server starts the creation of an adversarial sample against a pretrained CNN model written with the TensorFlow library.

Using TensorFlow utilizes the GPU of the server to quickly return results to the client, even if the client is a lesser-powered machine than the server. Evading samples are generated in real time (rather than returning a precomputed result), so the user may experience a delay if the attack they selected is particularly slow. The resulting adversarial sample and likelihoods for each class are displayed to the user through client-side code.

### 3.1 Design Decisions

In creating our system, we made several design decisions. Here, we present the reasoning behind the three largest system-level decisions we made: building ADVERSARIAL-PLAYGROUND as a web-based application, utilizing both client- and server-side code, and rendering images with the client rather than the server.

#### 3.1.1 Web-based Framework

The first question we asked was whether we wanted a web-capable framework or a desktop application. This

Figure 2: ADVERSARIAL-PLAYGROUND User Interface

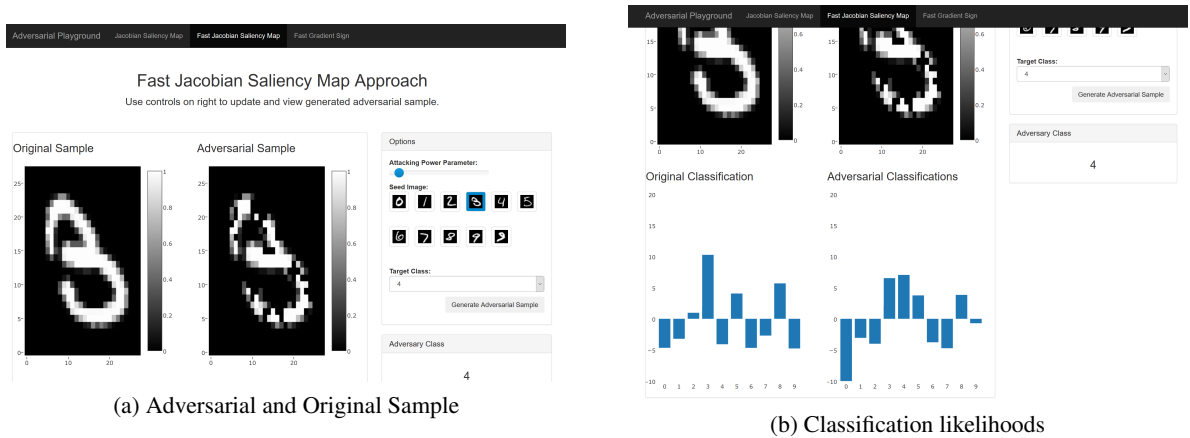

(a) Adversarial and Original Sample

(b) Classification likelihoods

was a fairly simple design choice, as a web-based interface enables a large number of users to utilize the application without requiring an installation process on each computer. By eliminating an installation step, we encourage potential users who may be only casually interested in adversarial machine learning to explore what it is. This supports the pedagogical goals of the software package.

### 3.1.2 Client-Server with Python back-end

Beyond a web-based framework, we needed to decide if ADVERSARIAL-PLAYGROUND would be client-based or if it would utilize server-side capabilities as well. The TensorFlow Playground was written entirely in JavaScript and other client-side technologies, allowing a lightweight server to host the service for many users. However, adversarial samples are usually generated on larger, deeper networks than those created by users of TensorFlow Playground, and this makes a JavaScript-only approach prohibitively slow.

Instead, we chose to use a GPU-enabled server running Python with TensorFlow to generate the adversarial samples. In addition to a speed advantage, this allows our baseline CNN model to be easily replaced with other TensorFlow graphs. Adding new attack strategies is simpler, too, as TensorFlow has been used in many research papers on evasion strategies.

### 3.1.3 Client-side rendering

Once we decided to generate the samples on the server-side, it was tempting to generate the output images on the server as well. This leads us to the third design decision we made: client-side rendering of MNIST images and likelihood plots. In our prototype, we used server-side rendering of these images using the Python library matplotlib, then loaded the image from the client directly.

However, this approach suffered from multiple disadvantages. Using only the default matplotlib utilities, we would have to write every generated image to disk and serve the unique URL to the user; this uses large amounts of disk, depending on the number of users. Furthermore, each image was around 20 kilobytes; waiting for the image to transfer added to the latency experienced by the user.

Fortunately, client-side rendering of images using the JavaScript library Plotly.JS resolved both of these issues. Rather than transferring an entire PNG image with thousands of pixels, we send only the pixel values for the $28 \times 28$ MNIST images and the 10 values for classification likelihoods within the POST response to the client. This reduced the amount of data we transferred on each request which, in turn, significantly reduced the time needed for each update. Additionally, this eliminates the need for storing any produced image files on the server. As a further advantage, the Plotly framework allows interactive charts that display individual pixel values on mouseover, providing users with more detailed feedback.

## 3.2 MNIST Dataset

ADVERSARIAL-PLAYGROUND uses the popular MNIST "handwritten digits" dataset for visualizing evasion attacks. This dataset contains 70,000 images of handwritten digits (0 through 9). Of these, 60,000 images are used as training data and the remaining 10,000 images are used for testing. Each sample is a $28 \times 28$ pixel, 8-bit grayscale image. Users of our system are presented with a collection of multiple seed images, selected from each of the 10 classes in the testing set[3].

We decided to only support the MNIST dataset for our visualization, as it is a common dataset used in evasion discussions. Many adversarial machine learning papers work with some form of image data; since this is easy to visually process, we thought it would be best to work with one of the MNIST, CIFAR, or ImageNet datasets. However, both ImageNet and CIFAR are much higher-dimensional than MNIST, which means generating adversarial samples is a more time-intensive process; the low-dimensionality of MNIST samples results in very fast generation of evasive samples. To provide the user with a low-latency yet representative experience, we used MNIST.

### 3.3 Software Manual

We released all project code on GitHub in the interest of providing a high-quality, easy-to-use software package to demonstrate evasion attacks. This section explains how to set up and use the package.

**Setup** Although minimal, the package requires a computer with Python 3.5, TensorFlow 1.0 (or higher), the standard `SciPy` stack, and the Python package `Flask`. We have tested the code on Windows, Linux, and Mac operating systems.

To install, clone the GitHub repository and install the prerequisites via `pip3 -r install requirements.txt`. A pre-trained MNIST model is already stored in the GitHub repository; all that is needed to start the webapp is to run `python3 run.py`. Once the app is started, it will run on `localhost:9000`.

**Usage** To use the application, the user selects a model from the navigation bar at the top of the webapp. On the right-hand pane, the user sets the attacking strength the algorithm using the slider, selects a seed image, and (if applicable) a target class. (Figure 2a at right.) Selecting a seed image immediately loads the image to the left-hand display.

After setting the parameters, the user clicks "Update Model." This runs the adversarial algorithm in real-time to attempt generating an evasive sample. The sample is displayed in the primary pane to the left of the controls (Figure 2a at left). This sample is fed through the classifier, and then the likelihoods are normalized and displayed in bar charts below the samples (Figure 2b). Finally, the classification of generated sample is displayed below the controls at right.

---

[3]A short script is in the project GitHub repository for generating new collections of seed images if the user runs the webapp locally.

## 4 Faster approach to JSMA

At the core of the ADVERSARIAL-PLAYGROUND is a set of pre-implemented attack models. It was important to present the user with the choice between targeted and untargeted approaches, as well as a choice between models utilizing different norms. As such, we implemented saliency-map based approaches (one directly from Papernot, et al. [8] and a faster one of our own development) for targeted methods and lightly modified the `cleverhans` implementation of the Fast Gradient Sign Method (FGS) for untargeted attacks.

As the FGS method is nearly identical to that found in `cleverhans`, we encourage the reader to consider [4] for implementation details. In the next two sections, we will review the details of the Jacobian Saliency Map Approach from the work of Papernot, et al., [8] and our improvement, *Fast Jacobian Saliency Map Apriori*.

### 4.1 Jacobian Saliency Map Approach (JSMA)

The Jacobian Saliency Map Approach (JSMA) adjusts the starting input to maintain similarity based on the $L^0$. Applied to the MNIST model, the approach is as follows:

1. Compute the forward derivative of the classifier, $\nabla F(X)$.

2. Use the saliency map of the sample to determine two best pixels to adjust.

3. Modify the two pixels and update the current sample.

4. Repeat until adversarial sample is misclassified or while the sample and the seed image differ by at most $\Upsilon$ (a tuneable threshold).

The first and last steps are fairly inexpensive in terms of time; the largest computational difficulty from using the saliency map to determine which pixels to adjust. In their original paper, Papernot et al. used Algorithm 1 for this selection process [8].

The key disadvantage of JSMA is that it must consider all pairs $(p,q)$ of possible feature indices (see Algorithm 1). Thus, the loop must perform $\Theta(|\Gamma|^2)$ iterations, where $|\Gamma|$ is the available feature size of each sample. When working on high-dimensional data, this becomes prohibitively expensive. By using a heuristic approximation of the JSMA algorithm, we achieve a faster runtime with comparable accuracy.

**Algorithm 1** Papernot's Saliency Map Feature Selection

$\nabla \mathbf{F}(\mathbf{X})$ is the forward derivative, $\Gamma$ the features still in the search space, and $t$ the target class

**Input:** $\nabla \mathbf{F}(\mathbf{X}), \Gamma, t$
1: **for** each pair $(p,q) \in \Gamma^2$, $p \neq q$ **do**
2: $\quad \alpha = \sum_{i=p,q} \frac{\partial \mathbf{F}_t(\mathbf{X})}{\partial \mathbf{X}_i}$
3: $\quad \beta = \sum_{i=p,q} \sum_{j \neq t} \frac{\partial \mathbf{F}_j(\mathbf{X})}{\partial \mathbf{X}_i}$
4: $\quad$ **if** $\alpha < 0$ and $\beta > 0$ and $-\alpha \times \beta >$ max **then**
5: $\quad\quad p_1, p_2 \leftarrow p, q$
6: $\quad\quad max \leftarrow -\alpha \times \beta$
7: $\quad$ **end if**
8: **end for**
9: **return** $p_1, p_2$

**Algorithm 2** Fast Jacobian Saliency Map Apriori Feature Selection

$\nabla \mathbf{F}(\mathbf{X})$, $\Gamma$, and $t$ as in Algorithm 1, $k$ is a small constant

**Input:** $\nabla \mathbf{F}(\mathbf{X}), \Gamma, t, k$
1: $K = \arg\text{top}_{p \in \Gamma} \left( -\frac{\partial \mathbf{F}_t(\mathbf{X})}{\partial \mathbf{X}_p}; k \right)$
2: **for** each pair $(p,q) \in K \times \Gamma$, $p \neq q$ **do**
3: $\quad \alpha = \sum_{i=p,q} \frac{\partial \mathbf{F}_t(\mathbf{X})}{\partial \mathbf{X}_i}$
4: $\quad \beta = \sum_{i=p,q} \sum_{j \neq t} \frac{\partial \mathbf{F}_j(\mathbf{X})}{\partial \mathbf{X}_i}$
5: $\quad$ **if** $\alpha < 0$ and $\beta > 0$ and $-\alpha \times \beta >$ max **then**
6: $\quad\quad p_1, p_2 \leftarrow p, q$
7: $\quad\quad max \leftarrow -\alpha \times \beta$
8: $\quad$ **end if**
9: **end for**
10: **return** $p_1, p_2$

## 4.2 Fast Jacobian Saliency Map Apriori (FJSMA)

Our improved JSMA is inspired by the Apriori algorithm used in frequent set mining. The Apriori algorithm is a fast, greedy, "bottom-up" approach to determining item sets with minimal support [1]. It achieves its speed through *a priori* elimination of certain suboptimal item sets. Similarly, our algorithm eliminates some $(p,q)$ pairs through *a priori* knowledge about the sample.

Instead of exhaustively considering each feature pair $(p,q)$, we rank the elements in the feature set $\Gamma$ by the value of the Jacobian at that coordinate. (This is the contribution each element makes to $\alpha$ in Algorithm 1.) We then force the choice of $p$ to be from the best $k$ such features and select $q$ from the entire feature set with $p$ removed. Since this choice of $p$ means its contribution to $\alpha$ is large, it is likely the product $-\alpha \times \beta$ will also be large.

Thus, we omit a large number of the $(p,q)$ feature pairs through *a priori* knowledge derived from our heuristic. This alternative to Algorithm 1 is shown in Algorithm 2. If we denote $K = \arg\text{top}_{p \in \Gamma} \left( -\frac{\partial \mathbf{F}_j(\mathbf{X})}{\partial \mathbf{X}_i}; k \right)$, where $\arg\text{top}_{x \in A}(f(x); k)$ is the set consisting of the top $k$ elements in $A$ as ranked by $f$, then the loop in our Fast Jacobian Saliency Map Apriori (FJSMA) selection routine runs in $\Theta(|K| \cdot |\Gamma|)$ time, where $|K| \ll |\Gamma|$. Since determining the top $k$ features can be done in $\Theta(|\Gamma|)$ time, this is a net improvement in asymptotic terms.

## 4.3 Experimental Results

The improved speed of FJSMA when compared to JSMA is especially beneficial in the real-time environment of ADVERSARIAL-PLAYGROUND. Low-latency generation of adversarial inputs provides a much better user experience for this case.

However, while the theory suggests that FJSMA will run faster than JSMA and generate evading samples at a similar rate, we must verify this with experimentation. We compare our FJSMA implementation (using a variety of values for $k$ and $\Upsilon$) to the implementation of JSMA found in Cleverhans [4] as of April 5, 2017.

We compared both evasion algorithms using the MNIST dataset and TensorFlow tutorial implementation of a MNIST-specific CNN. After training the CNN network on the MNIST training set, we ran both evasion attacks on the 10000-sample MNIST testing. To evaluate the performance of each approach, we compared the *evasion rate* — the percentage of seed images that were successfully converted into evasive samples. This is the standard metric used to evaluate evasion algorithms. We want to verify that our FJSMA algorithm is faster than JSMA and evades the classifier at a similar rate.

We determined the evasion rate for a range of values of the $\Upsilon$ parameter. With the FJSMA algorithm, we also varied the value of $k$ to be used; to represent the performance for arbitrarily-sized feature sets, we set $k$ in this experiment to be a percentage of the feature set size. Intuitively, this $k$ value is a control on how tight of an approximation FJSMA is to JSMA; as $k$ grows larger, we should expect the performance of the two approaches to converge to each other.

Results of this experiment are summarized in Figure 3a. The `cleverhans` JSMA and the new FJSMA attack perform similarly for all tested values of $\Upsilon$ and $k$, with larger values of $k$ increasing the evasion rate. Curiously, for $k \geq 20\%$, our implementation of FJSMA outperforms that of `cleverhans`; this is likely due to implementation details[4].

---
[4]The original preprint reported different results; at that time, we relied on high-level Python constructs for much of the algorithm. Now, we use lower-level `numpy` functions, resulting in a much faster runtime.

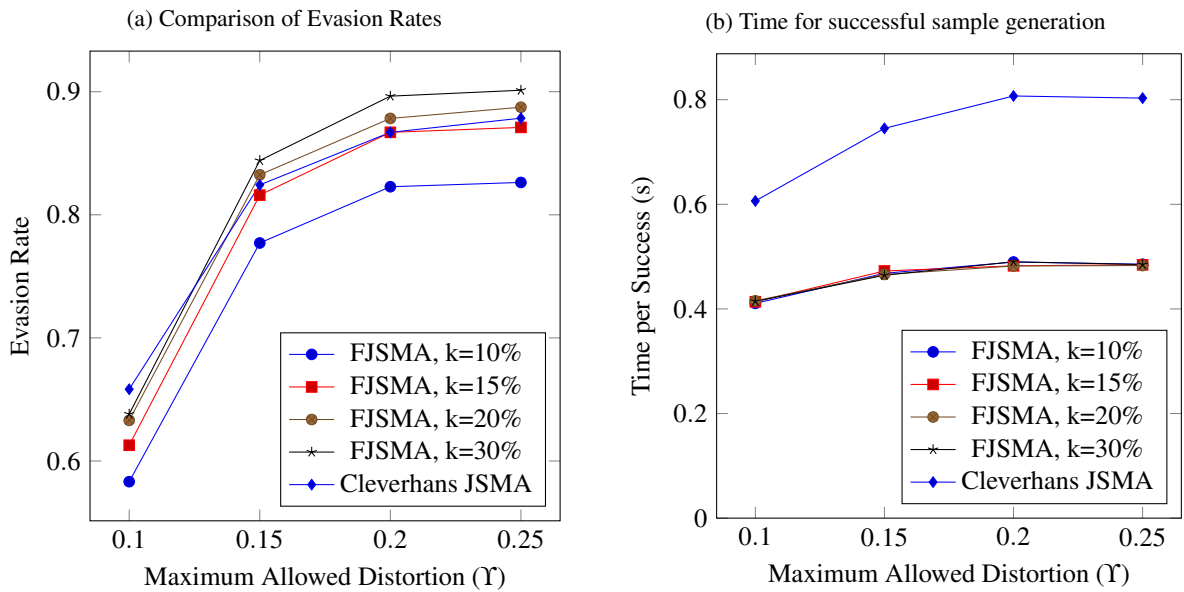

Figure 3: Experiment Results

In addition to evasion rate, we measured the "wall clock" time needed for successfully generating evasive samples. The average time to form an evasive sample from the original, benign sample is summarized in Figure 3b. We see that our FJSMA approach greatly improves upon the speed of a similarly written JSMA. However, it seems that varying the value of *k* does not produce a significant variation in runtime per sample. We conjecture this results from the small size of the feature space in MNIST and that searching 30% of the feature space likely does not dominate the runtime not large.

## 5 Discussion and Future Work

The study of evasion attacks on machine learning models is a rapidly growing field. In this paper, we present a web-based tool for visualizing the performance of evasion algorithms for deep neural networks. This helps both researchers and students alike to understand and compare the impact of adversarial examples against DNNs. Furthermore, we provide an improvement to the Jacobian Saliency Map Approach (JSMA) originally developed by Papernot, et al. [8]. This improvement, which we call the Fast Jacobian Saliency Map Apriori (FJSMA) approach, uses an *a priori* heuristic to reduce the search space significantly. FJSMA achieves a significant improvement in speed, while maintaining essentially the same evasion rate — an important advantage for visualization.

A straightforward extension of this work is to increase the variety of supported evasion methods. For example, including the new attacks based on $L^0$, $L^2$, and $L^\infty$ norms from Carlini and Wagner's recent paper [3] would be a good step in comparing the performance of multiple evasion strategies.

However, expansion in this manner presents an additional issue of latency. To generate evading samples "on-demand," the adversarial algorithm must run quickly; these other algorithms take much longer to execute than those we selected. A possible fix is to pre-compute results for slower methods with a selection of hyperparameter values, rather than allowing any value for hyperparameters. Another approach is to attempt applying the same greedy approximation technique used in FJSMA to the other attack algorithms; however, this may not always be possible.

Another direction for development is to provide a choice of classifiers and datasets. Allowing the user to select from CIFAR, ImageNet, and MNIST data would highlight the similarities and differences between how a single attack method deals with different data. Similarly, providing the user with a choice of multiple pre-trained models — possibly hardened against attack through adversarial training — would help distinguish artifacts of model choice from the behavior of the attack. These two extensions would help users more fully understand the behavior of an adversarial algorithm.